\documentclass[a4paper]{jpconf}
\usepackage{graphicx}
\usepackage{xcolor}
\usepackage{amsmath}  
\usepackage{amsfonts} 

\begin{document}
\title{A home-lab to study uncertainties using smartphone sensors and
determine the optimal number of  measurements}

\author{Martín Monteiro$^{1}$, Cecilia Stari$^2$, Arturo C. Mart{\'i}$^2$ }

\address{$^1$ Universidad ORT Uruguay, Montevideo, Uruguay}
\address{$^2$ Instituto de F\'{i}sica, Universidad de la
  Rep\'{u}blica, Montevideo, Uruguay}

\ead{marti@fisica.edu.uy}

\begin{abstract}
We present a home-lab experimental activity, successfully proposed to
our students during covid19 pandemic, based on
\textit{state-of-the-art} technologies to teach error analysis and
uncertainties to science and engineering students. In the last decade
the appearance of smartphones considerably affected our daily
life. Thanks to their built-in sensors, this revolution has impacted
in many areas and, in particular, the educational field. Here we show
how to use smartphone sensors to teach fundamental concepts for
science students such as any measurement is useless unless a
confidence interval is specified or how to determine if a result
agrees with a model, or to discern a new phenomenon from others
already known.  We explain how to obtain and analyse experimental
fluctuations and discuss in relation with the Gaussian distribution.
In another application we show how to determine the optimal number of
measurements as a function of the standard error and the digital
resolution of a given sensor.
\end{abstract}

\section{Introduction}

Statistical analysis is usually introduced in introductory laboratory
courses in first years of science studies
\cite{taylor1997introduction,hughes2010measurements}. The typical
approaches are based on performing manually repetitive observations,
for example, measuring a time interval with a stopwatch under
identical conditions. This kind of measurement gives slightly
different results due to the fact in the real world there exist
statistical fluctuations. The measurements obtained are usually
examined from the statistical viewpoint plotting histograms,
calculating mean values and standard deviations and, eventually,
compared with those expected from a known distribution, typically a
normal distribution. Commonly these experiments are tedious and
require a lot of time to be able to perform an adequate number of
measurements to perform the statistical analysis (see for example the
classical experiments proposed in
Refs. \cite{fernando1976experiment,mathieson1970student}).

The use of smartphone sensors has been incorporated in Physics
laboratories in all the fields, see for example
\cite{vieyrafive,monteiro2022resource,Lahme2024}.  Worth mentioning
characteristics of modern smartphones are the ability to measure
simultaneously with more than one sensor \cite{monteiro2019physics} or
supplement the data with video analysis \cite{monteiro2021allies}.
A very important advantage is that students usually have smartphones
and can therefore conduct experiments at home (see for example
\cite{monteiro2018bottle,Monteiro2022home,torriente2023rlc}).  In almost all the experiments, the focus is
on the mean values reported by the sensors while the fluctuations play
an annoying role.  Recently, it has been proposed a novel approach to
study uncertainties and errors in Physics laboratories taking the
advantages provides by smartphone sensors \cite{monteiro2021using}. In
this work, based on data recorded with the sensors present in many
mobile devices, we propose a set of experiments that can be performed
at home to teach error and uncertainties. In particular we discuss the
role of the digital resolution of the sensors and how to relate this
quantify to the statistical magnitudes to determine the optimal number
of repetitive measurements.

\section{A first approach to deal with statistics}

In this work we focus on the teaching of statistical uncertainties
which due to a multitude of causes are inherent to all physical
measurements \cite{taylor1997introduction}.  We assume that in a given
experiment an observation is repeated $N$ times under identical
conditions obtaining different results $x_i$, with $i=1,.. ,N$. It can
be shown that the best representative or \textit{estimate} of the set
of values is given by the mean value $\overline{x}$ defined as
\begin{equation}
 \overline{x}= \frac{1}{N} \sum_{i=1}^N x_i.
 \end{equation}
The deviation with respect to the mean value is identified with
$\epsilon_i= x_i - \overline{x}$. It can be shown that the mean value
defined as above minimizes the sum of the squared
deviations. Intuitively, it can be regarded as the
\textit{center-of-mass} of the set of the observations or equivalent
to the value \textit{closest} to all the other values.  In statistical
uncertainties it is of interest to quantify the dispersion of the values
around the mean value or, informally, the\textit{ width of the cloud}.
The standard deviation defined as
\begin{equation}
  \sigma = \sqrt{\frac{1}{N-1}\sum_{i=1}^N (x_i - \overline{x})^2}
\end{equation}
can be seen as a measure of this dispersion. If the number of
observations, $N$, is large enough, $\sigma$ it is characteristic of
the set of all the possible observations and does not depend on the
specific set of observations.  In practice, the uncertainty in the
determination of a physical magnitude depends on the number of
repeated measurements we have done.

\begin{figure}[h]
\begin{center}
\includegraphics[width=0.29\textwidth]{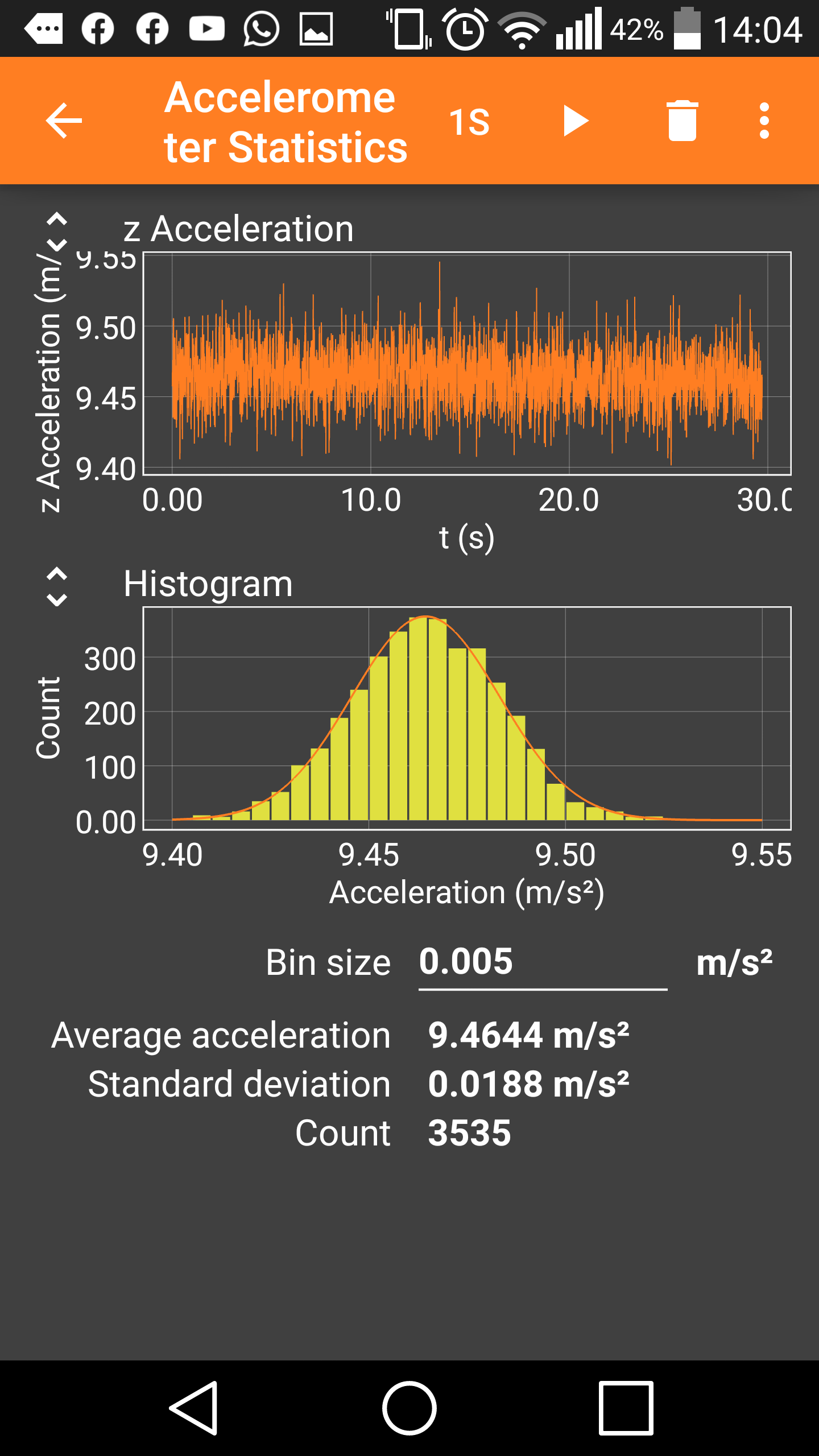}
\caption{Phyphox screenshot of the vertical component of the
  accelerometer while the smartphone is resting on a table.}
\label{fig:screenshot}
\end{center}
\end{figure}

The standard deviation, if $N$ is large enough, is characteristic of
the set of all the possible observations whereas the standard error,
or standard deviation of the mean, defined
as $$\sigma(\bar{a_z}) = \sigma_{a_z}/\sqrt{N}$$ represents the margin
of uncertainty of the mean value obtained in a particular set of
measurements \cite{taylor1997introduction}. The result of a specific
measurement is usually expressed as $\overline{a}_z \pm
\sigma(\bar{a_z})$ representing the best estimate and 68\% confidence
in that value.

It is worth highlighting that the standard deviation is related to the
degree to which an observation deviated from the mean value whereas
the standard error is an estimate of the uncertainty of the mean
value.  In a practical situation the standard uncertainty depends on
the number of measurements taken with $N^{-1/2}$. Then, given a set of
$N$ measures the standard deviation gives an idea of the dispersion of
an ideal set of infinite measures while the standard uncertainty
represents the uncertainty of our particular set of measurements. This
uncertainty can be reduced by increasing the number of measurements,
however, the square-root implies that this reduction is relatively
slow.

\section{Errors and uncertainties using smartphone sensors}

Mobile devices such as smartphones or tablets include several sensors
(accelerometer, magnetometer, ambient light sensor, among others) that
can be used as an alternative proposal to deal with statistical
distributions. The unavoidable fluctuations present in the several
sensors, so annoying in any measurement, can be used, however,
favourably, to illustrate basic concepts of statistical treatment of
errors and uncertainties \cite{monteiro2021using}. Using these
sensors, it is possible to acquire hundreds or thousands of repeated
measurements of a physical magnitude in a few seconds and analyse them
in the mobile device or transferred to a personal computer or analysed
in the cloud.

To access data registered by the sensors it is necessary to use an
special piece of software, or \textit{app}. There are many suitable
options available in the digital stores. In this proposal we mainly
use the Phyphox app \cite{staacks_2018} which stands out because of
its friendly interface and the possibility to access to experiments
programmed or proposed by other users.  As an example we show in
Fig.~\ref{fig:screenshot} the statistics of the vertical component of
the accelerometer value while the smartphone is simply resting on a
table. The graph on the top is the temporal evolution of the values
during an interval of about 20 $s$. We can observe that the values
oscillate around the gravitational acceleration value. The
corresponding histogram is plotted on the bottom panel where a
Gaussian fit is overlapped. Mean value, standard deviation, bin size
and number of measurements are also indicated on the bottom.

\begin{figure}[h]
\begin{center}
\includegraphics[width=0.78\textwidth]{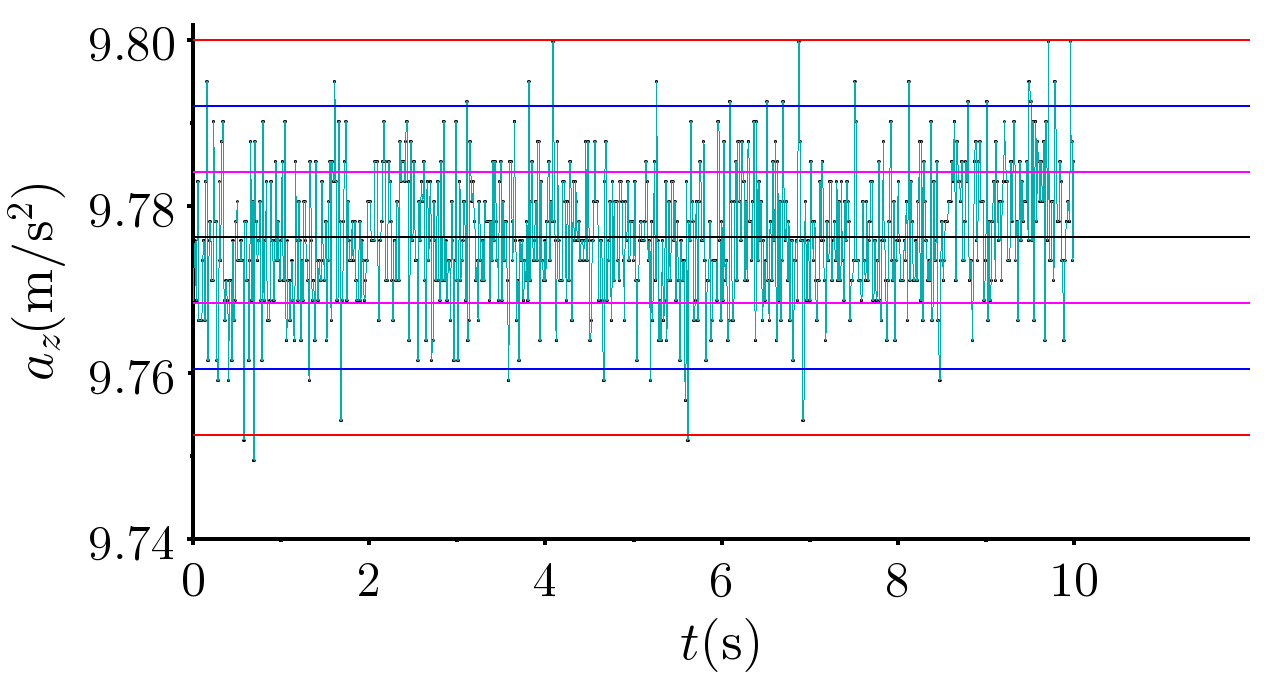}
\includegraphics[width=0.78\textwidth]{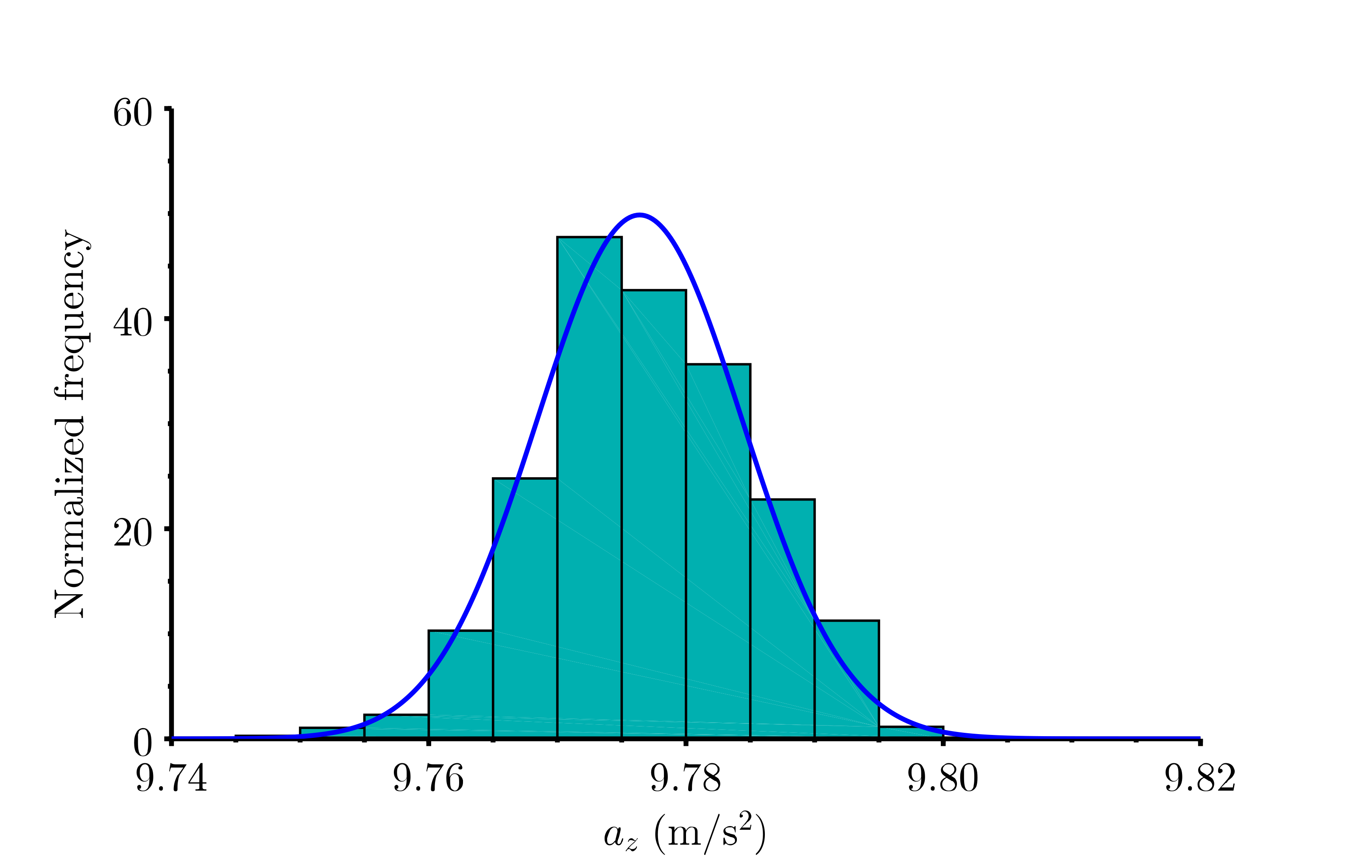}
\caption{Measurements of $z$ acceleration as a function of time for
  the smartphone held on the hand (top panel) and the histogram with a
  Gaussian fit (bottom panel).  }
\label{fig:z}
\end{center}
\end{figure}

As mentioned before, sensor data can be analysed in the smartphone or
transferred to a computer. Next, we show in Fig.~\ref{fig:z} the
temporal evolution of the vertical acceleration (top panel). We also
included in this graph horizontal lines at values $\overline{x} \pm
\sigma$, $\overline{x} \pm 2 \sigma$ and $\overline{x} \pm 3
\sigma$. These interval can be used to count the number of values in
each interval and compare with the well-known expected percentages
according to a Gaussian distribution.  In the bottom panel we show the
histogram and the corresponding Gaussian fit using the mean value and
standard deviation of the experimental data.

\begin{figure}[h]
\begin{center}
\includegraphics[width=0.8\textwidth]{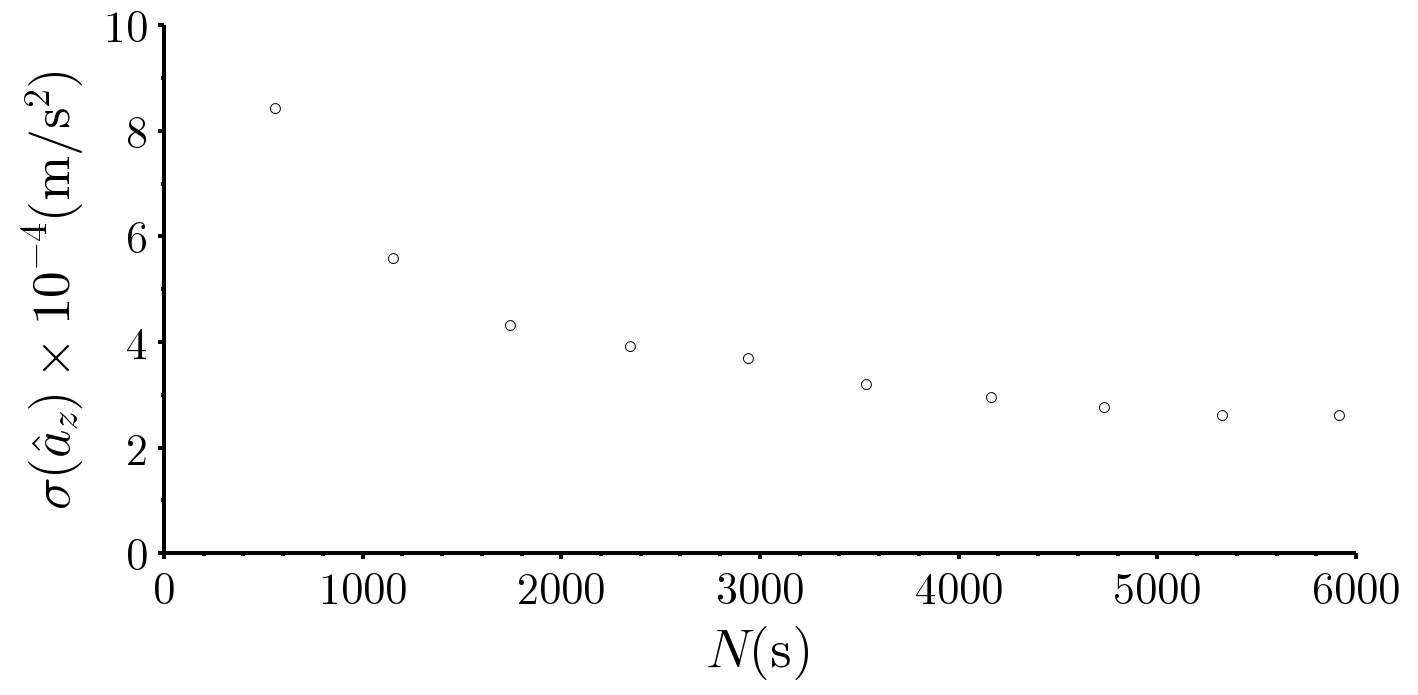}
\caption{Standard error as a function of the number of repetitive
  measurements. We notice that the standard uncertainty slowly
  decreases as we increase $N$.}
\label{fig:n}
\end{center}
\end{figure}

An interesting point is to consider, taking advantage of the
capability to quickly register data, different number of repetitive
measurements $N$ for the same experimental setup.  In
Table.~\ref{tabsigmaN} we report results of the mean value, standard
deviation and standard error for different length of these
experimental series. This information is also plotted in
Fig.~\ref{fig:n} where we show the standard uncertainty as a function
of $N$. As mentioned above, it is clear from that data that
$\sigma_{a_z}$ is nearly constant and, as a consequence,
$\sigma(\bar{a_z})$ is proportional to $N^{-1/2}$. According to our
experience, this graph results more convincing to the students than
theoretical explanations.

\begin{table}
 \begin{center}
\begin{tabular}{|c|c|c|c|} \hline 
$N$ & $\overline{x}$ & $\sigma$ & $\sigma_x$\\ \hline 
563 & 9,493 & 0,020 & 0,00085\\ 
1156 & 9,487 & 0,019 & 0,00054\\
1746 & 9,478 & 0,018 & 0,00044\\
2348 & 9,469 & 0,019 & 0,00039\\
2941 & 9,466 & 0,020 & 0,00036\\
3535 & 9,464 & 0,019 & 0,00032\\
4166 & 9,462 & 0,019 & 0,00029\\
4733 & 9,464 & 0,019 & 0,00027\\
5327 & 9,465 & 0,019 & 0,00026\\
5919 & 9,464 & 0,020 & 0,00026 \\ \hline
\end{tabular}
\caption{\label{tabsigmaN} Mean value, standard deviation and standard
  error as a function of the number of measurements corresponding to
  several experiments under identical conditions.}
\end{center}
\end{table}

\begin{figure}[tbp]
\begin{center}
\includegraphics[width=0.85\textwidth]{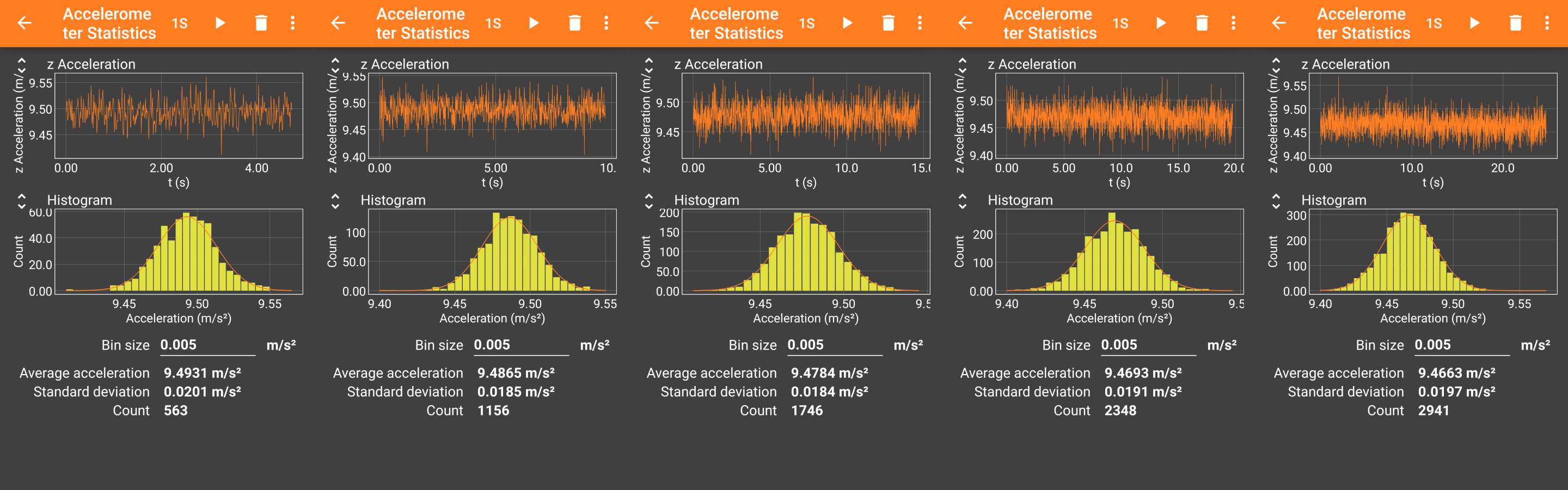}
\includegraphics[width=0.85\textwidth]{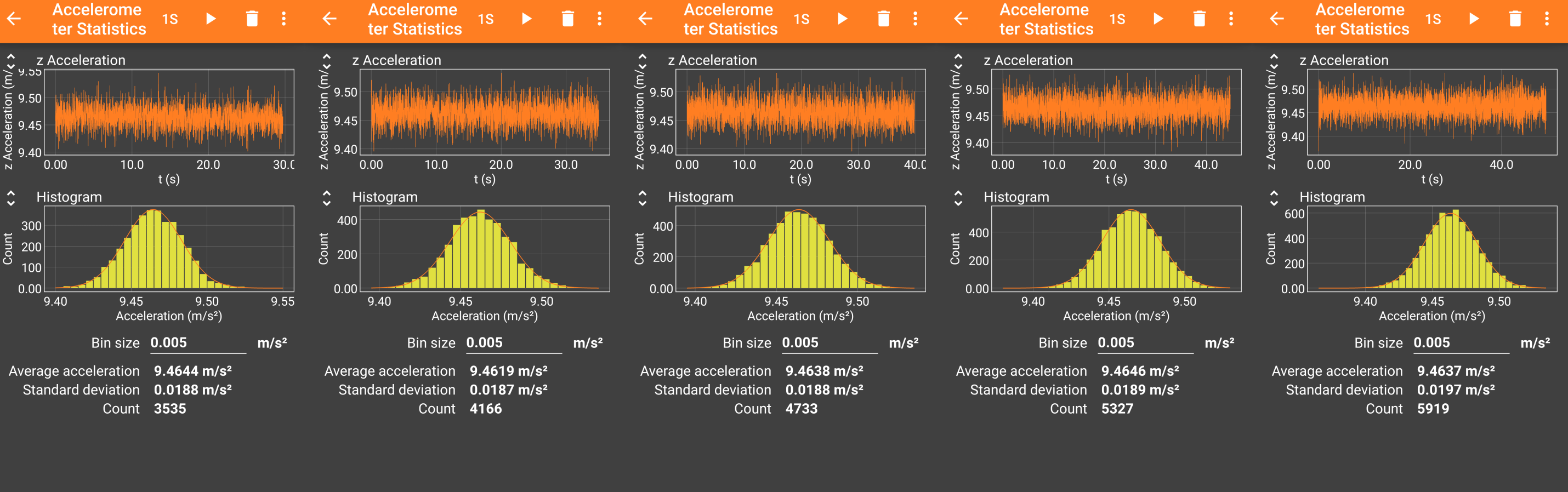}
\caption{Phyphox screenshots for different values of $N$. We observe
  as the number of repetitive measurements is increased the histograms
  curve results closer to the expected Gaussian fit.}
\label{fig:screenshots}
\end{center}
\end{figure}

\begin{figure}[tbp]
\begin{center}
\includegraphics[width=0.320\textwidth]{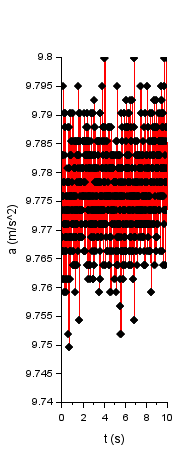}
\caption{A zoom of the temporal evolution of the $z$ component of the
  acceleration where we can observe the discrete nature of the values
  reported by the sensor.}
\label{fig:discrete}
\end{center}
\end{figure}

\section{Digital resolution and optimal number of measurements}

A closer look at the temporal evolution of the accelerometer data
reveals that the distribution in the vertical axis is not continuous
but instead present discrete values separated by uniform
intervals. The minimum interval corresponds to the digital resolution
of the sensor.  In Fig.~\ref{fig:discrete} we present a zoom of the
$z$ component of the acceleration in which we can better appreciate
the discrete distribution. The digital resolution can be estimate from
this graph to be of the order of $10^{-3}$ m/s$^2$.

The choice of $N$ in a specific experiment is a delicate question.
Indeed, if we could repeat the measurements infinite times the
standard uncertainty would vanish and we could achieve a perfect
knowledge of the best estimate.  However, as the decrease of the
standard uncertainty with the number of observations is slow, it is
impractical to increment this number excessively. A common criterion
is to take a number of measurements, often referred as the
\textit{optimal number of measurements}, $N_{\mathrm{opt}}$, such that
the statistical uncertainty is of the same order as the systematic (or
type B) errors. Here, in the absence of other sources of systematic
errors, the standard uncertainty should be of the same order as the
resolution of the digital instrument: $\sigma(\bar{a_z}) =
\sigma_{a_z} /\sqrt{N_{opt}} \sim \delta$. In the experiment depicted
in Table~\ref{tabsigmaN} with a LG G3, the resolution is
$\delta=0.0012$ m/s$^2$, therefore $N_{\mathrm{opt}} \sim 250.$ It is
worth emphasizing that this estimate is optimistic and must be
examined in case of having other sources of systematic errors.

\section{Other applications}

\begin{figure}[tbp]
\begin{center}
\includegraphics[width=0.35\textwidth]{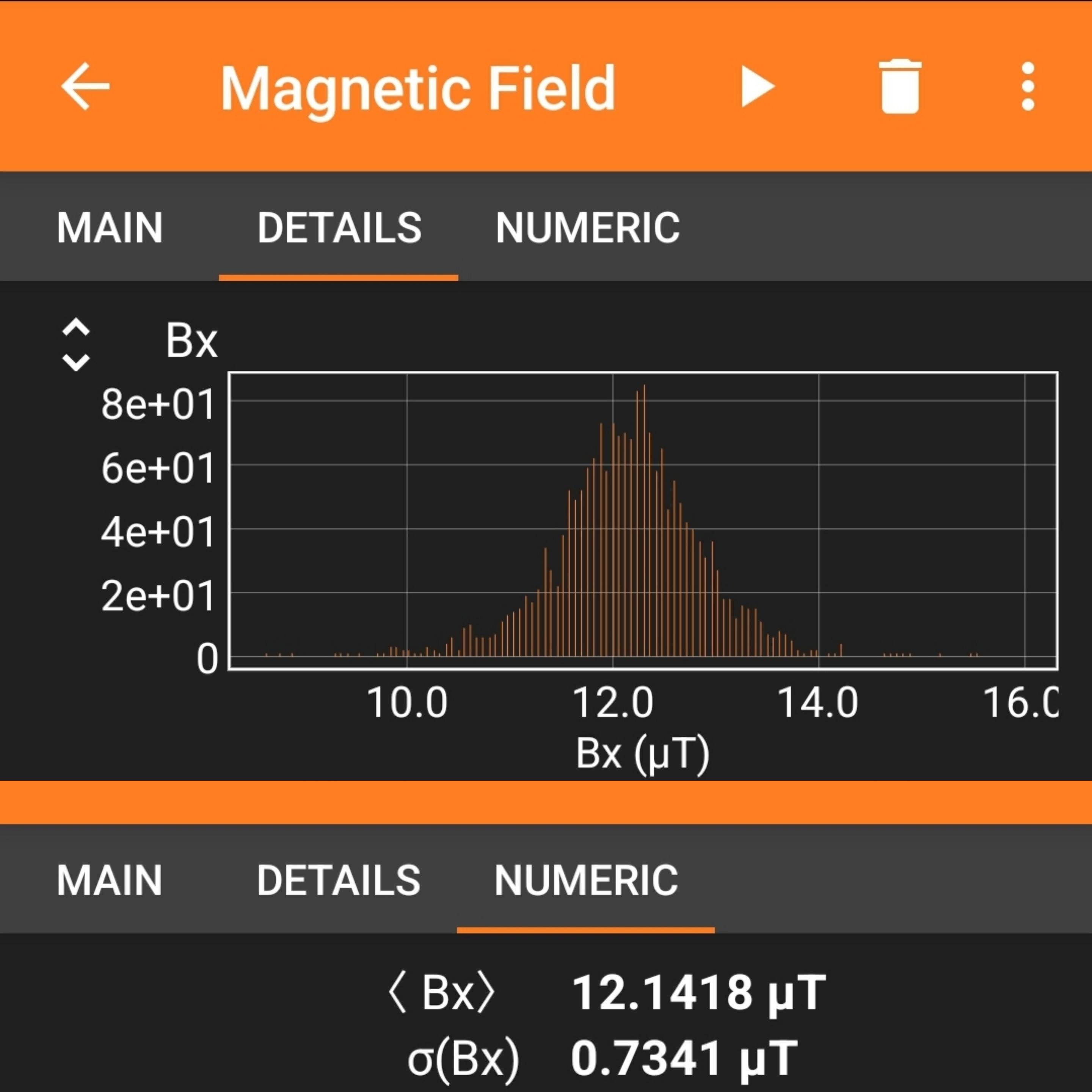}
\caption{Phyphox screenshot showing an histogram and the Gaussian fit
  of the magnetometer data. Note that this data set in considerable
  noisier than that obtained with the accelerometer.}
\label{fig:mag}
\end{center}
\end{figure}

This set of activities is just a sample of all that can be done by
linking sensors and statistical fluctuations.  It is possible to
consider variations of the mean values or of the standard deviation or
width of the distributions \cite{monteiro2021using}. This quantity is
linked to the intensity of the fluctuations.  As possible applications
we can mention proposing a challenge to the students to take the
picture or selfie with the lowest level of fluctuations. Another
application would be to evaluate the comfort of a means of transport,
car, bus, plane or even the state of a traffic road
\cite{harikrishnan2017vehicle}.  Another experiment (not recommended
by the authors) is to study the fluctuations of the gait of a
pedestrian as a function of the alcohol beverage intake, similarly to
Ref.~\cite{zaki2020study}.  It is also possible to analyse the
statistical character of other magnitudes different from the
acceleration. To show another possibility, Fig.~\ref{fig:mag}
reproduces a screenshot of a similar experiment using, in this case,
the magnetometer. This analysis can be useful when considering outdoor
experiments dealing with magnetic field as proposed in
Ref. \cite{monteiro2020magnetic}.
 
\section{Closing remarks}

The COVID19 pandemic strongly affected experimental activities in
introductory physics courses almost everywhere in the world. This
forced us to reformulate traditional instructional laboratories to
make it possible to conduct them in students' homes or even outdoors.
As a positive by-product of these restrictions we proposed this home
laboratory to experiment with statistical errors and uncertainties and
involves fundamental concepts related such as mean values, standard
distributions, histograms and fluctuations.  The main tool used are
the built-in sensors embedded in smartphones that are widely available
to students.  These devices allow to obtain in a few minutes an
important number of repetitive measurements. In this work we focus
precisely on determining the optimal number of these measurements and
link it to the digital resolution of the sensors. Before that, the
introductory laboratory to statistical analysis was carried out by
means of repetitive  measurements.

The evaluation of this activity has been very positive and we can
mention some didactic advantages that we have been able to appreciate.
On the one hand, the activity can be carried out in an agile way,
without waiting times and with the possibility of having the results
immediately.  These aspects make it possible to allocate more time to
the discussion of the relevant aspects.  We have found that this
clearly helps students gain a significant understanding of the
physical meaning of histograms, standard deviation, and other
statistical tools.  Another notable advantage that we have appreciated
is that the use of their own devices increases the autonomy of the
students, who can apply what they have learned in various situations
outside the laboratory and of which we have taken advantage by
inviting the students to observe what happens when they are traveling
by car or bus. Finally, we can mention that in general we have been
able to notice that the students show more interest and commitment
when carrying out these activities with the sensors than when we
carried out the traditional laboratory.  These experiments could
contribute to motivating students and showing them the necessity of
considering uncertainty analysis.  Several possible extensions related
to non-normal statistics can be considered, such as Poisson
distribution \cite{mathieson1970student}, distribution of maxima,
Chauvenet criterion \cite{limb2017inefficacy}, or Benford's law
\cite{bradley2009benford}.

\section{Acknowledgments} We acknowledge financial support from grant
Fisica Nolineal (ID 722), Grupos I+D CSIC 2018 (UdelaR,
Uruguay) and PEDECIBA (UdelaR, MEC, Uruguay).

\section*{References}

\bibliography{/home/arturo/Dropbox/bibtex/mybib}

\providecommand{\newblock}{}
\begin{thebibliography}{10}
\expandafter\ifx\csname url\endcsname\relax
  \def\url#1{{\tt #1}}\fi
\expandafter\ifx\csname urlprefix\endcsname\relax\def\urlprefix{URL }\fi
\providecommand{\eprint}[2][]{\url{#2}}

\bibitem{taylor1997introduction}
Taylor J 1997 {\em Introduction to error analysis, the study of uncertainties
  in physical measurements\/} (University Science Books)

\bibitem{hughes2010measurements}
Hughes I and Hase T 2010 {\em Measurements and their uncertainties: a practical
  guide to modern error analysis\/} (Oxford University Press)

\bibitem{fernando1976experiment}
Fernando P~C~B 1976 {\em American Journal of Physics\/} {\bf 44} 460--463
  \urlprefix\url{https://doi.org/10.1119/1.10177}

\bibitem{mathieson1970student}
Mathieson E and Harris T~J 1970 {\em American Journal of Physics\/} {\bf 38}
  1261--1262 \urlprefix\url{https://doi.org/10.1119/1.1976040}

\bibitem{vieyrafive}
Vieyra R, Vieyra C, Jeanjacquot P, Marti A and Monteiro M 2015 {\em The Science
  Teacher\/} {\bf 82} 32--40

\bibitem{monteiro2022resource}
Monteiro M and Martí A~C 2022 {\em American Journal of Physics\/} {\bf 90}
  328--343 ISSN 0002-9505 (\textit{Preprint}
  \eprint{https://pubs.aip.org/aapt/ajp/article-pdf/90/5/328/16438261/328\_1\_online.pdf})
  \urlprefix\url{https://doi.org/10.1119/5.0073317}

\bibitem{Lahme2024}
Lahme S~Z, Klein P and Müller A 2024 {\em Journal of Physics: Conference
  Series\/} {\bf 2693} 012008
  \urlprefix\url{https://dx.doi.org/10.1088/1742-6596/2693/1/012008}

\bibitem{monteiro2019physics}
Monteiro M, Stari C, Cabeza C and Martí A~C 2019 {\em Journal of Physics:
  Conference Series\/} {\bf 1287} 012058
  \urlprefix\url{https://dx.doi.org/10.1088/1742-6596/1287/1/012058}

\bibitem{monteiro2021allies}
Monteiro M, Cabeza C, Stari C and Marti A~C 2021 {\em Journal of Physics:
  Conference Series\/} {\bf 1929} 012038
  \urlprefix\url{https://doi.org/10.1088/1742-6596/1929/1/012038}

\bibitem{monteiro2018bottle}
Monteiro M, Stari C, Cabeza C and Marti A~C 2018 {\em The Physics Teacher\/}
  {\bf 56} 644--645 (\textit{Preprint}
  \eprint{https://doi.org/10.1119/1.5080589})
  \urlprefix\url{https://doi.org/10.1119/1.5080589}

\bibitem{Monteiro2022home}
Monteiro M, Stari C and Mart{\'\i} A~C 2022 {\em Physics Education\/} {\bf 58}
  013003

\bibitem{torriente2023rlc}
Torriente-Garc{\'\i}a I, Mart{\'\i} A~C, Monteiro M, Stari C, Castro-Palacio
  J~C and Monsoriu J~A 2023 {\em Physics Education\/} {\bf 59} 015016

\bibitem{monteiro2021using}
Monteiro M, Stari C, Cabeza C and Martí A~C 2021 {\em American Journal of
  Physics\/} {\bf 89} 477--481 ISSN 0002-9505
  \urlprefix\url{https://doi.org/10.1119/10.0002906}

\bibitem{staacks_2018}
Staacks S, H\"{u}tz S, Heinke H and Stampfer C 2018 {\em Physics Education\/}
  {\bf 53} 045009

\bibitem{harikrishnan2017vehicle}
Harikrishnan P and Gopi V~P 2017 {\em IEEE Sensors Journal\/} {\bf 17}
  5192--5197

\bibitem{zaki2020study}
Zaki T~H~M, Sahrim M, Jamaludin J, Balakrishnan S~R, Asbulah L~H and Hussin F~S
  2020 The study of drunken abnormal human gait recognition using accelerometer
  and gyroscope sensors in mobile application {\em 16th IEEE International
  Colloquium on Signal Processing \& Its Applications (CSPA)\/} (IEEE) pp
  151--156

\bibitem{monteiro2020magnetic}
Monteiro M, Organtini G and Mart{\'\i} A~C 2020 {\em The Physics Teacher\/}
  {\bf 58} 600--601

\bibitem{limb2017inefficacy}
Limb B~J, Work D~G, Hodson J and Smith B~L 2017 {\em Journal of Fluids
  Engineering\/} {\bf 139}

\bibitem{bradley2009benford}
Bradley J~R and Farnsworth D~L 2009 {\em Teaching Statistics\/} {\bf 31} 2--6

\end{thebibliography}

\end{document}